\documentclass[
aps,
prresearch,
twocolumn,
superscriptaddress,
amsmath,
amssymb
]{revtex4-2}

\usepackage{amsmath,amssymb,amsfonts}
\usepackage{physics}
\usepackage{algorithmic}
\usepackage{graphicx}
\usepackage{textcomp}
\usepackage{orcidlink}
\usepackage{cleveref}
\usepackage[normalem]{ulem}

\newcommand{\bs}[1]{\boldsymbol{#1}}
\usepackage{xcolor}
\def\BibTeX{{\rm B\kern-.05em{\sc i\kern-.025em b}\kern-.08em
    T\kern-.1667em\lower.7ex\hbox{E}\kern-.125emX}}

\newtheorem{thm}{Theorem}[section]

\newtheorem{cnj}[thm]{Conjecture}
\newtheorem{dfn}[thm]{Definition}
\newtheorem{rem}[thm]{Remark}
\newtheorem{exa}[thm]{Example}

\begin{document}

\title{Beyond Quantum Advantage: Improved Classical Algorithms for the Binary Paint Shop Problem}

\author{Mark Goh}
\affiliation{Institute for Frontier Materials on Earth and in Space, German Aerospace Center (DLR), 51170 Cologne, Germany}
\affiliation{Institute for Theoretical Physics, University of Cologne, Cologne, Germany}

\author{Lara Caroline Pereira dos Santos}
\affiliation{Institute for Frontier Materials on Earth and in Space, German Aerospace Center (DLR), 51170 Cologne, Germany}

\author{Thorge Mueller}
\affiliation{Institute for Software Technology, Department High-Performance Computing, German Aerospace Center (DLR), 51170 Cologne, Germany}

\author{Matthias Sperl}
\affiliation{Institute for Frontier Materials on Earth and in Space, German Aerospace Center (DLR), 51170 Cologne, Germany}
\affiliation{Institute for Theoretical Physics, University of Cologne, Cologne, Germany}

\begin{abstract}
The binary paint shop problem (BPSP) is an APX-hard optimization problem in which, given $n$ car models that occur twice in a sequence of length $2n$, the objective is to find a colouring sequence such that each car model pair is painted differently while minimizing the number of times the paint is swapped along the sequence.
A recent classical heuristic, known as the recursive star greedy (RSG) algorithm, is conjectured to achieve an expected paint swap ratio of $0.361$, thereby outperforming the Quantum Approximate Optimization Algorithm (QAOA) with circuit depth $p=7$. Since the performance of the QAOA with logarithmic circuit depth is instance independent, the average paint swap-ratio is upper-bounded by the QAOA. We provide an improved upper-bound of the BPSP by extending the QAOA to depth $p=17$, outputting an expected paint swap ratio of $0.334$ via an exact computation while numerical extrapolation suggests a further reduction to a value of $0.295$. To provide hardware-relevant comparisons, we additionally implement the BPSP on a D-Wave Quantum Annealer Advantage 2, obtaining a minimum paint swap ratio of $0.329$. Given that the QAOA with logarithmic circuit depth does not exhibit a quantum advantage for sparse optimization problems such as the BPSP, this implies the existence of a classical algorithm that outperforms both the RSG algorithm and logarithmic depth QAOA. We provide numerical evidence that the Mean-Field Approximate Optimization Algorithm (MF-AOA) is one such algorithm, yielding a paint swap ratio of approximately $0.280$ beating all known classical and quantum algorithms for the BPSP.
\end{abstract}

\maketitle

\section{Introduction}
Quantum computing is expected to offer a computational speedup for certain optimization problems. For instance, Grover's algorithm for unstructured search \cite{Grover} and Shor's algorithm for factoring \cite{Shor1} both achieve an asymptotic speedup against the best classical algorithm for their respective problem. However, these algorithms require fault-tolerant quantum computers which are out of reach in the current Noisy Intermediate-Scale Quantum (NISQ) \cite{Preskill} era of quantum computing. As a result, there is a growing interest in quantum algorithms that can demonstrate a quantum advantage with the current hardware. 

A leading candidate is the Quantum Approximate Optimization Algorithm (QAOA) \cite{QAOA_paper}. Inspired by quantum adiabatic computing \cite{Adiabatic}, the QAOA is a heuristic algorithm that, given $p$ layers, provides an approximate solution for combinatorial optimization problems. Under a change of basis, one can express these problems as an Ising model with spin variables $z\in \{-1,1\}$, where the cost function can be expressed as
\begin{align}
    H(z)= \sum_{i}h_i z_i + \sum_{i<j} J_{ij}z_i z_j ,
\end{align}
and $h_i$ and $J_{ij}$ are real-valued numbers with $h_i$ being the local magnetic field and $J_{ij}$ the coupling strength between spins $z_i$ and $z_j$.

Additionally, one can use quantum annealers to solve such combinatorial optimization problems. Quantum annealing (QA), as an open quantum adiabatic hardware system, has been made available by D-Wave and it is known to be an efficient QUBO and Ising solver for small problem sizes \cite{berwald2018,Albash2018,king2025,Pelofske2024,QA_MCPSP,Sandt2023}.

In a previous application of the QAOA on the binary paint shop problem (BPSP), it was found that the QAOA outperforms the recursive greedy algorithm (RSG), the best classical heuristic at the time of publication, requiring a depth of only $p=7$ \cite{QAOA_BPSP}. Since then, a new recursive star greedy algorithm was developed that is conjectured to outperform $p=7$ QAOA \cite{RSG_BPSP}. Furthermore, it has been recently argued that for sparse optimization problems such as the BPSP, there can be no quantum advantage offered by the QAOA with logarithmic circuit depth when compared to approximate message passing (AMP) algorithms \cite{chen2023localalgorithmsfailurelogdepth}. This implies that even if the QAOA outperforms the newly created recursive star greedy algorithm at constant depth, there exists a classical algorithm that would outperform the QAOA. Here we investigate the hypothesis that the Mean-Field Approximate Optimization Algorithm (MF-AOA), a classical algorithm inspired by the QAOA and developed in Ref.\ \cite{MFAOA}, is one such algorithm which could outperform logarithmic depth QAOA in the BPSP.

The paper is structured as follows: In \cref{sec:background}, we review the existing background covering the BPSP and the relevant algorithms of this paper. \Cref{sec:method} outlines the relevant methods used to evaluate the performance of the QAOA, QA and MF-AOA on the BPSP. We then provide the results of our paper in \cref{sec:results}, before discussing open questions and concluding remarks in \cref{sec:conclusion}.

\section{Background}
\label{sec:background}

\subsection{Binary Paint Shop Problem}

The paint shop problems are a set of various optimization problems in the automobile industry, where given a sequence of cars with a certain constraint on how to paint them, the number of times the paint colour is swapped must be minimized. The problem was first introduced in Ref.\ \cite{PSP}, where it was proven to be an NP-complete problem if either the number of colours or the number of car body types is unbounded. Loosely speaking, in decreasing order of difficulty, there are three main types of paint shop problems: i) the multi-car multi-colour paint shop problem, where we have multiple occurrences of each car body type with multiple paint colours, ii) the multi-car paint shop problem, where we restrict to only two colouring, and iii) the binary paint shop problem (BPSP), where each car type occurs only twice. This paper focuses on the BPSP, defined as the following:

\begin{dfn}(Binary Paint Shop Problem (BPSP))
\newline Let $A=\{a_1,a_2,\dots, a_n\}$ be the set of car body types. Let $S=(s_1,\dots, s_{2n})$ be a sequence of car instances such that each $a_i\in A$ appears exactly twice in $S$. Denote the two paint colours as the set $C=\{0,1\}$. The BPSP aims to find a colouring sequence $f=(f_1,\dots,f_{2n})\in \{0,1\}^{2n}$ that minimises the number of colour changes $\Delta_C$ between consecutive cars with the constraint that for each car body type, the two occurrences must be coloured differently (i.e.,\ for any car body type $a_i$, if $s_j=s_k=a_i$ for $j\not = k$, we have $f_j \not = f_k$). This can be expressed mathematically as
\begin{align}
    \begin{aligned}
        \mathrm{minimize }\; & \Delta_C=\sum_{i=1}^{2n-1} [f_i \not = f_{i+1}],\\
        \mathrm{subject}\text{ }\mathrm{to}\;  &f_i \not = f_j \quad \forall i,j: s_i=s_j, i\not = j,
    \end{aligned}
\end{align}
where $[\cdots]$ denotes the Iverson bracket.
\end{dfn}

\begin{exa}
Consider the following instance of an $n=3$ BPSP problem with the sequence
\begin{align*}
    S= (a_1,a_2,a_1,a_3,a_3,a_2).
\end{align*}
A valid solution corresponds to the colouring 
\begin{align*}
    f'=(0,0,1,1,0,1), \quad \Delta_C'=3.
\end{align*}
However, this is not optimal since we have
\begin{align*}
     f'=(0,1,1,1,0,0), \quad \Delta_C'=2.
\end{align*}
\end{exa}

The BPSP is an NP-complete problem in its decision form \cite{PSP} and APX-hard as an optimization problem \cite{BPSP_APX}. Thus, unless P = NP, no polynomial-time algorithm can find an arbitrary approximate solution for all instances, let alone solve them exactly. In order to evaluate the performance of various algorithms, consider the following expression: for fixed $n$, let the normalised expectation of $\Delta_C$ over all possible sequences of length $2n$ be given by $\mathbb{E}\Delta_C/n$. Denote the average performance of an algorithm $\mathcal{A}$ on the BPSP be $\mathbb{E} \mathcal{A}(\Delta_C/n)$. The performance of a classical greedy algorithm and recursive greedy algorithm have been found in Ref. \cite{BPSP_greedy} to produce an expected colour change of
\begin{align*}
    \text{greedy }&: \lim_{n\rightarrow \infty}\mathbb{E} \mathcal{A}\left(\frac{\Delta_C}{n}\right)=0.5 + \mathcal{O}(1/n) ,\\
    \text{recursive greedy }&:  \lim_{n\rightarrow \infty} \mathbb{E}\mathcal{A}\left(\frac{\Delta_C}{n}\right)=0.4 + \mathcal{O}(1/n).
\end{align*}

This raises an interesting question. Namely, what is the value of $\mathbb{E} \Delta_C/n$ in the limit $n\rightarrow \infty$? It was conjectured in Ref. \cite{BPSP_con} that $\mathbb{E}\Delta_C/n=o(n)$ but this has been proven false in Ref.\ \cite{BPSP_lowerbound} where they proved the following theorem.
\begin{thm}[Modified Theorem 1.4 of Ref.\ \cite{BPSP_lowerbound}]
The expected colour change $\mathbb{E}\Delta_C/n$ is lower bounded by 
\begin{align}
    2H^{-1}\left(\frac{1}{2}\right)-o(1)= 0.220\ldots -o(1),
\end{align}
where $H^{-1}$ is the inverse of the binary entropy function. In addition, it is upper bounded by $0.4+o(1)$.
\begin{rem}
The authors further noted that the lower bound can be slightly improved to 0.227, and they provided an outline of the proof.
\end{rem}
\end{thm}

The authors of Ref.\ \cite{RSG_BPSP} independently proved a weaker lower bound of 0.214 and slightly improved the upper bound to $0.4-\epsilon$ for sufficiently small $\epsilon>0$. In addition, they developed a new recursive star greedy algorithm that is conjectured to have an upper bound of $0.361$. Even if their conjecture is true, there is still a substantial gap between the upper and lower bound.

\subsection{BPSP as a QUBO}

The algorithm to convert any instance of the BPSP into an Ising problem (more precisely, a weighted MaxCut problem) was developed in Ref.\ \cite{QAOA_BPSP} and further made mathematically rigorous independently in Ref.\ \cite{RQAOA_BPSP} and Ref.\ \cite{GQAOA_BPSP}. For completeness, we describe the steps in transforming the BPSP into an Ising problem. Each car body type $i$ is associate with a single qubit $z_i$ with the sign of the spin corresponding to the colour assigned to the first instance of car body type in the sequence. Let $\Omega: S\rightarrow A$ be a map that labels each $s_i$ to an instance of a car body type in $A$. The couplings between the spins are calculated by going down the sequence $S$, with the coupling strength between the spin representation of adjacent cars $s_i$ and $s_{i+1}$ determined by the car body type and whether they have occurred in the sequence previously. If both car body types are distinct and $s_{i}$ and $s_{i+1}$ are both the first or second appearance of their car type, a coupling strength of $-1$ between the corresponding spins in the Ising formulation is added. If one of the car is the first occurrence of their car type and the other the second occurrence, a coupling strength of $+1$ is added between the corresponding spins. If both cars are of the same body type, a constant is added since we have $Z_iZ_i=1$ for all $i$. Finally, a factor of $1/2$ is added for each term so that the energy of the Ising formulation corresponds to the number of colour changes. Thus, for $n$ car body types, the BPSP-Ising Hamiltonian is given by
\begin{align} \label{eq:hamiltonian1}
    H_{\mathrm{BPSP}}&= \frac{1}{2}
    \sum_{i=1}^{2n-1} \left(
        J_{s_i s_{i+1}}' Z_{\Omega(s_i)}Z_{\Omega(s_{i+1})} + 1
    \right)\cr
    &=\frac{1}{2}
    \sum_{i=1}^{2n-1}J_{s_i s_{i+1}}' Z_{\Omega(s_i)}Z_{\Omega(s_{i+1})}
    + \frac{2n-1}{2},
\end{align}
where $J_{s_i s_{i+1}}'$ is the coupling strength assigned by the relation between adjacent cars $s_{i}s_{i+1}$ as defined above and $\Omega(s_i)$ the car body type of the $i$-th car in the sequence.

The equation above can be further simplified by converting the sum over adjacent cars in the sequence into a sum over car body types $A$. Doing so results in a simplified Hamiltonian given by
\begin{align} \label{eq:hamiltonian2}
    H_{\mathrm{BPSP}}= \frac{1}{2}
    \sum_{(i,j)\in E}J_{ij} Z_{i}Z_{j} + \frac{c'}{2} ,
\end{align}
where $J_{ij}$ is the effective coupling of the Ising spins corresponding to the $i$ and $j$ car body type, $E$ is the set of edges in the corresponding Ising graph, and $c'=2n-1+ \# \text{adjacent pairs of the same car type}$.
\begin{rem}
 Note that the matrix $J$ is a triangular matrix rather than a symmetric matrix which is typical for Ising problems due to the directed sequence/graph. If one wishes to use a symmetric formulation, then the factor of $1/2$ has to be changed to a factor of $1/4$.
\end{rem}

As we are interested in averaging over all sequences $S$ in the limit $n\rightarrow \infty$ to find the average colour change $\mathbb{E}\Delta_C/n$, we note that $\lim_{n\rightarrow \infty}\mathbb{E} c'/n=2$. Furthermore, it has been shown previously that in the large $n$ limit, the coupling strength only takes on the values $J_{ij}\in \{-1,+1\}$, with -1 occurring with a probability of $2/3$ and +1 with a probability of $1/3$, and that the underlying graph approaches that of a 4-regular tree \cite{QAOA_BPSP}. Thus, we have
\begin{align}
    \label{eq:BPSP_Ising}
   \lim_{n\rightarrow \infty} \frac{H_{\mathrm{BPSP}}}{n}= 
    \frac{1}{2n}\sum_{(i,j)\in E}J_{ij} Z_{i}Z_{j} + 1,
\end{align}
where $|E|=2n$.

\subsection{Quantum Approximate Optimization Algorithm }

The Quantum Approximate Optimization Algorithm (QAOA) \cite{QAOA_paper} is a quantum variational algorithm for finding approximate solutions to combinatorial optimization problems. The problem is mapped into an Ising formulation typically encoded as a Quadratic Unconstrained Binary Optimization (QUBO) problem or more generally a Higher order/Polynomial Unconstrained Binary Optimization (HUBO/PUBO) problem. Focusing on the BPSP, the Hamiltonian can be expressed as
\begin{align}
    \hat{H}_c = \frac{1}{2}\sum_{i,j}J_{ij} \hat{Z}_i\hat{Z}_j ,
\end{align}
where $\hat{Z}_i$ is the Pauli $Z$ operator acting on the $i$-th qubit and $J_{ij}$ is the coupling strength between the $i$-th and $j$-th qubits. The constant term has been dropped since it adds only an overall phase factor to the quantum state and a shift in the energy.

The unitary operators of the QAOA are defined as
\begin{align}
    \hat{U}(\beta)&= \exp{-i \beta \sum_j \hat{X}_j},\\
    \hat{U}(\gamma)&= \exp{-i \gamma \hat{H}_c},
\end{align}
where $\gamma,\beta\in \mathbb{R}$ and $\hat{X}_i$ is the Pauli $X$ operator acting on the $i$-th qubit.

The QAOA ansatz state is then defined by the number of layers $p$ and $2p$ variational parameters
\begin{align}
    \ket{\bs{\gamma},\bs{\beta}}= \prod_{i}^p  \hat{U}(\beta_i) \hat{U}(\gamma_i)\ket{+}^{\otimes N},
\end{align}
where $\bs{\gamma}=(\gamma_1,\dots,\gamma_p)\in \mathbb{R}^{p}$,  $\bs{\beta}=(\beta_1,\dots,\beta_p)\in \mathbb{R}^p$, and $\ket{+}$ the uniform superposition state.

The objective function to be optimized in the QAOA is typically the expected energy 
\begin{align}
    E(\bs{\gamma},\bs{\beta})=\expval{H_c}{\bs{\gamma},\bs{\beta}}.
\end{align}

\subsection{Mean-Field Approximate Optimization Algorithm}

The Mean-Field Approximate Optimization Algorithm (MF-AOA) \cite{MFAOA} is a quantum-inspired classical limit of the QAOA, where the time evolution of the quantum spins is replaced by classical spin dynamics via the mean-field approximation.

For a concrete example of how the MF-AOA works, consider an Ising model with spins $\bs{Z}\in \{-1,1\}^N$ with a local magnetic field $h_i\in \mathbb{R}$:
\begin{align}
     H_{h,q}(\bs{Z})= \sum_{i=1}^N h_i Z_i + \sum_{j,k} J_{jk} Z_j Z_k,
\end{align}
where we excluded the normalising constant for ease of notation.

As a classical algorithm inspired by the QAOA, one similarly uses a problem Hamiltonian and a driving Hamiltonian of the following form:

\begin{subequations}
\begin{equation}
    H_P = -H_{h,q}(\bs{Z}) ,
    \end{equation}
    \begin{equation}
    H_D = -\sum_{i=1}^N \Delta_i X_i,\quad \Delta_i>0,
\end{equation}
\end{subequations}
where the positivity of $\Delta_i$ ensures that the ground state of $H_D$ is given by $\ket{+}^{\otimes N}$.

The classical spin vector is defined as
\begin{align}
    \bs{n}_i(t) &= (n_i^x(t),n_i^y(t),n_i^z(t))\cr
    &=(\text{Tr}[\rho X_i(t)],\text{Tr}[\rho Y_i(t)],\text{Tr}[\rho Z_i(t)]),
\end{align}
with $\rho$ being the density matrix in the mean-field approximation $\rho=\otimes_{i=1}^N \rho_i$.

In the mean-field approximation, the total Hamiltonian has the form
\begin{align}
    H(t) = -\gamma(t) \sum_{i=1}^N \left[h_i + \sum_{j} J_{ij} n_j^z \right]n_i^z - \beta(t) \sum_{i=1}^N \Delta_i n_i^x,
\end{align}
where $\beta(t)$ and $\gamma(t)$ are piecewise-constant functions. 

Define also the effective magnetization as
\begin{align}
    \label{eq:eff_mag}
     m_{i}(t)=h_i + \sum_{1 \le j\le N } J_{ij} n_{j}^z(t).
\end{align}

The dynamics of this system boils down to solving the $3N$ differential equations for each spin in its effective magnetic field
\begin{align}
    \partial_t \bs{n}(t) = \bs{n}(t) \times \bs{B}(t),
\end{align}
with $B_i(t) = \langle 2(1-s(t))  ,0 ,2 s(t) m_i(t)\rangle$.

\begin{rem}
For the BPSP, without an external field (i.e.,\ $ \forall i\; h_i =0$) the effective magnetization remains 0. In order for the algorithm to work, one could fix the last spin of the state, i.e.,\ $n_N^z=1$ in order to break the $\mathbb{Z}_2$ symmetry.
\end{rem}

The piecewise solution to the mean-field Hamiltonian is thus given by
\begin{align}
    \label{eq:evolutions}
        n_i (p) = \prod_{t=1}^p V_i^D(t) V_i^P(t) n_i (0) ,
    \end{align}
    with   
    \begin{subequations}
    \label{eq:evolutions_2}
        \begin{align}
            V_i^D(t) &=\begin{pmatrix}
                1 & 0 & 0\\
                0 & \cos{(2\Delta_i \beta_t)} & \sin{(2\Delta_i \beta_t)}\\
                0 & -\sin{(2\Delta_i \beta_t)} & \cos{(2\Delta_i \beta_t)}
            \end{pmatrix},\\
            V_i^P(t) &=\begin{pmatrix}
                \cos{( 2m_{i,t-1}\gamma_t)}  & \sin{(2 m_{i,t-1}\gamma_t)} & 0\\
                -\sin{( 2m_{i,t-1}\gamma_t)} & \cos{( 2m_{i,t-1}\gamma_t)}& 0\\
                0 &0 & 1
            \end{pmatrix}.
        \end{align}
    \end{subequations}

Since the QAOA is guaranteed to find the exact solution by reduction to the adiabatic algorithm, it suffices to take linear functions of $\gamma$ and $\beta$ as in Ref.\ \cite{Willsch_2020} for a sufficiently large number of time step $T$ to ensure the algorithm achieves a near optimal performance. 

We now explain the algorithm. Starting with $N$ classical spins in the $x$-axis, for each time step, it applies the mean-field evolution using \cref{eq:evolutions_2} and calculates the new effective magnetization using \cref{eq:eff_mag}. Once finished, it rounds off the $z$ component of each spin vector to obtain the bitstring $\sigma^* = \left( \text{sign}(n_1^z), \dots,\text{sign}(n_N^z)  \right)$.

\subsection{Related Work}

There have been a few other works studying the performance of QAOA variants on the BPSP. Ref.\ \cite{RQAOA_BPSP} examines the performance of Recursive-QAOA (RQAOA) on the BPSP and found that parameter transfer showed no noticeable reduction in solution quality for both QAOA and RQAOA. In addition, it was found that RQAOA outperformed the standard QAOA. Ref.\ \cite{GQAOA_BPSP} similarly studies the performance of RQAOA and eXpressive-QAOA (XQAOA) with a circuit depth $p=1$, where not only XQAOA outperformed RQAOA, but the authors also provided evidence that the performance of RQAOA diminishes as the problem size increases.

Regarding quantum annealing, Ref.\ \cite{QA_MCPSP} studies the application of QA on the multi-car paint shop problem, where it was found to exhibit an advantage on small problem instances but it loses any advantage in the asymptotic limit, performing comparably to a simple greedy algorithm. A recent work studied the performance of Decoded Quantum Interferometry (DQI) and benchmarked it against a state of the art classical algorithm concluding that it is unlikely for DQI to exhibit a quantum advantage on the BPSP \cite{dqi_bpsp}.

\section{Methods}
\label{sec:method}

\subsection{Application of the QAOA to the BPSP}
In the large $n$ limit, the BPSP is equivalent to the Max-2-XORSAT problem where the underlying graph is locally a signed 4-regular tree. It is known that up to logarithmic-depth, the performance of the QAOA on sparse instances of Max-2-XORSAT and MaxCUT are the same \cite{chen2023localalgorithmsfailurelogdepth,chou_et_al:LIPIcs.ICALP.2022.41}. An algorithm to evaluate the performance of the QAOA on $(D+1)$-regular trees was developed in Ref.\ \cite{QAOA_largegirth} for constant depth $p$, where the expectation value of the QAOA for a single edge is given by
\begin{align}
    \label{eq:QAOA_tree}
    \expval{Z_i Z_j}{\bs{\gamma},\bs{\beta}}=-\frac{2v_p(D,\bs{\gamma},\bs{\beta})}{\sqrt{D}},
\end{align}
with 
\begin{equation}
\begin{split}
    \nu_p(D,\bs{\gamma},\bs{\beta})
    &=
    \frac{i\sqrt{D}}{2}
    \sum_{\mathbf a,\mathbf b}
    a_0b_0\,
    f(\mathbf a)f(\mathbf b)\,
    H_D^{(p)}(\mathbf a)H_D^{(p)}(\mathbf b)
    \\
    &\hspace{1.0cm}\times
    \sin\!\left[
        \frac{
        \bs{\Gamma}\cdot(\mathbf a\mathbf b)
        }{\sqrt{D}}
    \right],
\end{split}
\label{eq:nu_finite_D_results}
\end{equation}
where \(\mathbf a\) and \(\mathbf b\) denote spin configurations, \(f\) contains the mixer matrix elements, and \(H_D^{(p)}\) is the depth-\(p\) branch message generated by the finite-\(D\) regular-tree recursion of Ref.~\cite{QAOA_largegirth}.

While the original algorithm has a time complexity of $\mathcal{O}(p16^p)$ and a memory complexity of $\mathcal{O}(4^p)$, a co-author of Ref.\ \cite{QAOA_largegirth} reformulated the same tree-QAOA algorithm as a tensor network applied on a regular tree to reduce the time complexity to $\mathcal{O}(4^p)$ \cite{QAOA_treetensor}, thereby achieving a quadratic speedup. Similar to the original algorithm of Ref.\ \cite{QAOA_largegirth}, because all branches of the tree are identical, only one branch must be contracted explicitly, while the remaining contributions are incorporated by raising the resulting tensor entry-wise to the appropriate branching power.

Since every edge in the graph is locally tree-like up to logarithmic depth, the validity of \cref{eq:QAOA_tree} can be extended from constant depth to logarithmic depth $p\sim \mathcal{O}(\log n)$ \cite{Goh1}. Thus, the performance of log-depth QAOA on \cref{eq:BPSP_Ising} can be simplified to
\begin{align}
\label{eq:QAOA_inf_p_n_BPSP}
   \lim_{p\rightarrow \infty} \lim_{n\rightarrow \infty} \mathbb{E}_J
    \expval{\frac{H_{\mathrm{BPSP}}}{n}}{\bs{\gamma},\bs{\beta}}
    =1- \frac{2\nu(3,\bs{\gamma},\bs{\beta})}{\sqrt{3}}.
\end{align}

We point out to the reader that in evaluating $\nu(3,\bs{\gamma},\bs{\beta})$, the unitary operator $\hat{U}(\bs{\gamma)}$ used in Ref.\ \cite{QAOA_largegirth} is slightly different. Instead of using the Problem Hamiltonian (here $H_{\mathrm{BPSP}}$), a scaled driving Hamiltonian with $ Z_i Z_j/\sqrt{D}$ rather than $Z_iZ_j/2$ is used. In our numerical simulation, we followed the convention of Ref.\ \cite{QAOA_largegirth} and thus, our values of $\gamma$ differ by a constant factor from the known literature values. The angles used in our simulation can be found in Appendix \ref{sec:angles}.

\subsection{Quantum Annealing}

Next, we implement the BPSP via quantum annealing using the Ising formulations in \cref{eq:hamiltonian1} and \cref{eq:hamiltonian2}. We apply the problem to the D-Wave QA Advantage 2, system 2, located in the central European region. For every problem size $n$, we solve 1000 randomly generated instances, each instance with 100 QA read outs. The QA annealing time is set to 100 $\mu s$. The results do not significantly change once these parameters (i.e., the number of read outs and the annealing time) are increased past our selected values. We then calculate the average cost, or average paint swap ratio, and the standard deviation normalized by $n$. 

\subsection{Simulation of MF-AOA}

We perform a numerical simulation of the MF-AOA modifying some of the codes that were developed in the original paper \cite{MFAOA_code}. To estimate the algorithm's performance in the large $n$ limit, we randomly generated 1000 instances of the BPSP for each problem size $n$ up to $n=10000$. Note that given a problem with size $n$, there is the question of deciding the number of time-step (aka.\ the runtime of the algorithm) $T$. One could, in principle, set $T$ to be arbitrarily large but we found that when we set $T\gg n$, the results were worse than simply setting $T=n$. For example, for $n=2000$ instances, 4 different runs of the MF-AOA were done with $T\in \{1000,2000,5000,10000\}$. We found that most instances had the best results using $T=2000$ while only a few had slightly better results with $T=5000$. Thus, simply setting $T$ to be arbitrarily large does not seem to be the optimal solution. In order to simplify the analysis, we shall set $T=\max(1000,n)$.

\section{Results and Discussions}
\label{sec:results}
\subsection{QAOA}
\label{subsec:qaoa_result}
For the BPSP, the normalized expected number of paint changes is according to equation \ref{eq:QAOA_tree} and \ref{eq:QAOA_inf_p_n_BPSP}. Thus, the values reported below are not obtained by simulating a finite BPSP instance, but by contracting the infinite-tree light cone of a representative edge. The simulations of Ref.~\cite{QAOA_BPSP} are used only as an independent validation at the depths for which they are available. 

\begin{figure}[t]
    \centering
    \includegraphics[width=\columnwidth]
    {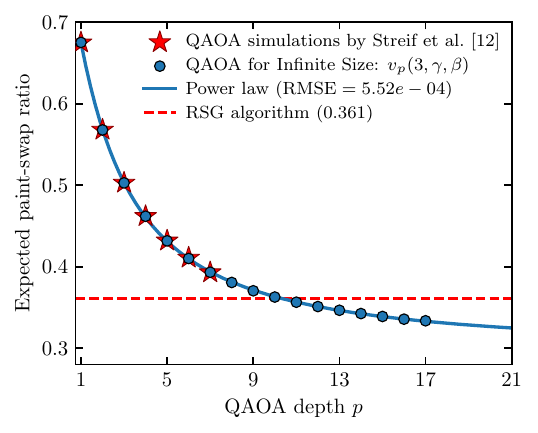}
    \caption{
    Expected paint-swap ratio of the BPSP as a function of the QAOA depth. Blue circles show the large-girth recursion results up to \(p=17\). The parameters for \(p\leq8\) were obtained by direct optimization, whereas those for \(p\geq9\) were generated by extrapolating the smooth lower-depth schedules and subsequently evaluated with the finite-\(D\) recursion. Red stars denote the QAOA simulations of Ref.~\cite{QAOA_BPSP}. The solid blue curve is a least-squares fit to the power law in eq.~\eqref{eq:qaoa_power_fit}; its root-mean-square error is \(5.52\times10^{-4}\). The horizontal dashed line marks the conjectured performance of the recursive star-greedy algorithm, \(\rho_{\mathrm{RSG}}=0.361\), and the first QAOA depth with a performance below this value is \(p=11\).
    }
    \label{fig:bpsp_recursion}
\end{figure}

Using the tensor-network structure, we extend the BPSP recursion results to \(p=17\). The parameter schedules for \(p\leq 8\) are obtained by direct optimization of the recursion and are the same as those found in Ref.\cite{QAOA_BPSP,Wybo} up to a rescaling. For \(p\geq 9\), we extrapolate the smooth lower-depth schedules and evaluate the resulting parameters with the same finite-\(D\) recursion, without an additional high-dimensional optimization. The complete parameter sets are listed in Appendix~\ref{sec:angles}. For \(p\leq 7\), the recursion values agree with the explicit QAOA simulations of Ref.~\cite{QAOA_BPSP} to the quoted precision; see Fig.~\ref{fig:bpsp_recursion} and Table~\ref{tab:comparison}. This agreement validates both the large-girth reduction and the angle-convention conversion. At larger depth, QAOA outperforms the recursive star-greedy algorithm at \(p=11\).

We estimate an upper bound of the performance of the QAOA using $\lim_{p\rightarrow \infty}v_p(3,\bs{\gamma},\bs{\beta})$, which is equivalent to the performance at logarithmic depth \cite{Goh1}. Since we are unable to evaluate this term explicitly, we fit the curve obtained using $v_p(3,\bs{\gamma},\bs{\beta})$ up to depth $p=17$ to get a numerical estimate of the upper-bound.

We use a power law fit for the QAOA's performance as a similar fit was done in Ref.\  \cite{QAOA_SK_evidence} for the QAOA's performance when applied to the Sherrington--Kirkpatrick model with good accuracy. In addition, several other works have found a power-law behaviour in the scaling exponent as in Ref.\ \cite{power_law,QAOA_SAT} which translates to a power-law behaviour in performance.

Assuming a power law fit to the function
    \begin{align}
    \label{eq:qaoa_power_fit}
     v_p(3,\gamma,\beta) \approx  d-\frac{a}{p^{b}+c},
    \end{align}
with parameters $(a,b,c,d)$. The fitted parameters are
\begin{subequations}
\begin{align}
a &= 1.12254 \pm 0.02286,\\
b &= 1.2182 \pm 0.0182,\\
c &= 2.4099 \pm 0.0827,\\
d &= 0.61065 \pm 0.00126.
\end{align}
\end{subequations}
with
\begin{equation}
    \operatorname{RMSE}
    =
    4.78 \times 10^{-4}.
\end{equation}
The small RMSE shows that the shifted power law accurately describes
the available finite-depth data.

Assuming that our power law fit holds, this suggests that at logarithmic depth, the performance of the QAOA for the binary paint shop problem is 
    \begin{align}
       \lim_{n\rightarrow \infty}\mathbb{E} \mathcal{A}\left(\Delta_C/n\right)= &1- \frac{2\nu(3,\bs{\gamma},\bs{\beta})}{\sqrt{3}} \approx 0.295 \quad  \text{with}\\
  0.293 \le  &1- \frac{2\nu (3,\bs{\gamma},\bs{\beta})}{\sqrt{3}} \le 0.296.
    \end{align}
Using the fitted value gives us an expected value of 0.295, while the range of values comes about from the uncertainty of the fit.

\subsection{Quantum Annealing}

Results for the quantum annealing solver are reported in \Cref{tab:QA-advantage2} and \Cref{fig:mfaoa_bpsp}. The best performance of the Advantage 2 solver is obtained with $n=40$, obtaining a minimum average paint swap ratio of $0.329$, which is then followed by an increase of the average paint swap ratio. We speculate that this loss of performance is primarily due to the current hardware limitations.

The minimum average paint swap ratio achieved by the QA solver is slightly lower than our corresponding values obtained with the QAOA up to depth $p=17$ (see \Cref{tab:comparison}, \Cref{fig:bpsp_recursion} and \Cref{fig:mfaoa_bpsp}). A scaling advantage of quantum annealing over simulated annealing and gate-based approaches has been suggested previously for certain optimization problems \cite{Albash2018,king2025,Pelofske2024}. However, this advantage has thus far remained limited to relatively small problem sizes, as for larger problem sizes, hybrid solutions are preferred \cite{QA_MCPSP,Sandt2023}. Our results are consistent with these previous observations, showing that the Advantage 2 solver provides competitive performance for small instances but does not exhibit a sustained advantage as the problem size increases.

\begin{table}[t]
    \centering
    \begin{tabular}{|c|c|c|}
       \hline
       $n$ & $\mathbb{E}\Delta_C/n$ & $\mathbf{s}(\Delta_C/n)$ \\
       \hline 
       10 & 0.3702 & 0.0903 \\
       \hline
       20 & 0.3411 & 0.0595 \\
       \hline
       30 & 0.3319 & 0.0487 \\
       \hline 
       40 & 0.3294 & 0.0412 \\
       \hline
       50 & 0.3308 & 0.0392 \\
       \hline 
       60 & 0.3351 & 0.0378 \\
       \hline
       70 & 0.3418 & 0.0366 \\
       \hline
       80 & 0.3516 & 0.0366 \\
       \hline 
       90 & 0.3602 & 0.0368 \\
       \hline
       100 & 0.3966 & 0.0414 \\
       \hline
    \end{tabular}
    \vspace{2mm}
    \caption{Results from D-Wave's QA Advantage 2 solver for 1000 randomly generated BPSP instances for each $n$.}
    \label{tab:QA-advantage2}
\end{table}

\subsection{MF-AOA}

Following the extrapolation of the QAOA's result in \Cref{subsec:qaoa_result}, we make a conjecture about the performance of the MF-AOA:
\begin{cnj}
\label{cnj:BPSP}
In the limit $n\rightarrow \infty$, AMP algorithms such as the MF-AOA achieve a performance of approximately
\begin{align}
   \lim_{n\rightarrow\infty} \mathbb{E} \Delta_C/n = 1-\frac{2\nu(3,\bs{\gamma},\bs{\beta})}{\sqrt{3}} \approx 0.295.
\end{align}
\end{cnj}
\begin{rem}
We use $\approx$ due to competing factors. First, the extrapolation is based on sub-optimal value of the QAOA's performance as the values for $p\ge 9$ we obtained through heuristic extrapolation and no further optimization were made. Thus, this contributes to an underestimation. However, an extrapolation of the QAOA's performance naturally leads to an overestimation as observed in Ref.\ \cite{QAOA_SK_evidence} where extrapolation of the QAOA's performance on the Sherrington--Kirkpatrick model with optimised circuit depth up to $p=80$ still leads to an overestimation of the true value but is contained within the range of uncertainty. As such, we cannot safely say whether our extrapolation is an over or under estimation but merely that the value should be within the ballpark range of 0.29.
\end{rem}

\begin{rem}
One of the authors has a preliminary proof that the MF-AOA can be understood as a type of AMP algorithm in their PhD thesis \cite{Goh_thesis}. However, the proof has some minor mistakes and has been corrected for an upcoming paper.
\end{rem}

\begin{figure}[t]
    \centering
    \includegraphics[width=\columnwidth]{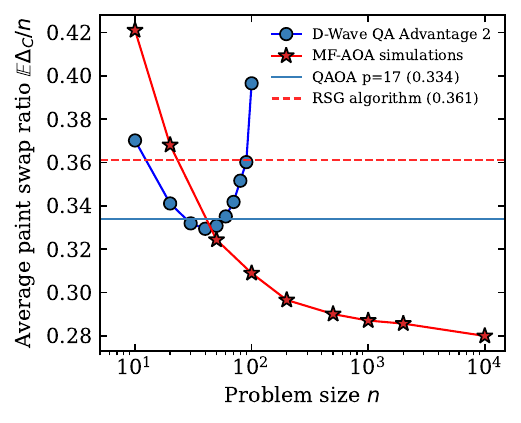}
    \caption{A semi-log plot of the average expected paint swap ratio achieved by the recursive star-greedy algorithm (RSG=0.361), the QAOA $p=17$ (0.334), the MF-AOA simulations and the D-Wave QA Advantage 2 Solver. Both the MF-AOA and D-Wave runs were averaged over 1000 randomly generated instances for each problem size $n$.}
    \label{fig:mfaoa_bpsp}
\end{figure}

We found that the MF-AOA has an expected paint swap ratio of $\mathbb{E}\Delta_C/n=0.27993$ with a standard deviation of $\mathbf{s}=0.001469$ for $n=10000$, outperforming all known classical heuristics, applications of the QAOA and QA to the BPSP. In general, both the expected paint swap ratio and the standard deviation decreases as $n$ increases. Further details can be found in \Cref{fig:mfaoa_bpsp} and \Cref{tab:mfaoa_bpsp}. Consistent with the findings of Ref. \cite{chen2023localalgorithmsfailurelogdepth}, we conjecture that AMP algorithms are optimal for average-case sparse optimization problems. 

The difficulty with proving conjecture \ref{cnj:BPSP} is that, to the best of the authors' knowledge, the performance guarantee of AMP algorithms is only well understood when, loosely speaking, one takes the large $n$ limit followed by setting the average degree of the graph to infinity (i.e.,\ $\lim_{D\rightarrow\infty}\lim_{n\rightarrow \infty}$). For the BPSP, the degree is fixed at $D=4$. The performance of AMP algorithms on MaxCut is known \cite{AMP_maxcut} to be bounded by
\begin{align}
    \frac{1}{2}+\Pi_2 \sqrt{\frac{1}{D}} + o_D\left(\frac{1}{\sqrt{D}}\right),
\end{align}
where $\Pi_2= 0.763\dots$ is the ground state energy of the Sherrington--Kirkpatrick model.

Converting the cut ratio to a paint swap ratio, this gives us a value of 
\begin{align}
    \mathbb{E} \Delta_C/n &= 1- \Pi_2 +  o_D\left(\frac{1}{\sqrt{D}}\right) \cr
    &= 0.2368\dots + o_D\left(\frac{1}{\sqrt{D}}\right).
\end{align}
For $D=4$, the sub-leading term is not suppressed and cannot be ignored leading to a correction that is difficult to quantify. 

Similarly, assuming that the bounds for MaxCut and Max-2-XORSAT (and hence BPSP) are similar, then using the known upper-bound of 0.8683 and lower-bound of 0.8333 for Maxcut on a 4-regular graph \cite{maxcut_bound}, the expected paint swap ratio would be lower-bounded by 0.2634 and upper-bounded by 0.3334. Our numerical results for the expected paint swap ratio of approximately 0.27993 fits within these bounds which gives us good reason to believe that the algorithm is near-optimal.

\begin{table}[t]
    \centering
    \begin{tabular}{|c|c|c|}
       \hline
       $n$  &  $\mathbb{E}\Delta_C/n$ & $\mathbf{s}(\Delta_C/n)$\\
       \hline 
       10 & 0.4211 & 0.1200\\
       \hline
        20 & 0.3681 & 0.0664\\
        \hline
        50 & 0.3243 & 0.0383\\
        \hline
        100 & 0.3089 & 0.0226\\
        \hline 
        200 & 0.2965 & 0.0147\\
        \hline
        500 & 0.2900 & 0.0089\\
        \hline 
        1000 & 0.2870 & 0.0055\\
        \hline
        2000 & 0.2856 & 0.0047\\
        \hline
        10000 & 0.2799 & 0.0014\\
        \hline 
    \end{tabular}
    \vspace{2mm}
    \caption{Results from the numerical simulations of the MF-AOA over 1000 randomly generated BPSP instances for each $n$.}
    \label{tab:mfaoa_bpsp}
\end{table}

\section{Conclusions}
\label{sec:conclusion}
In this paper, we have analyzed the performance of the QAOA, QA and the MF-AOA for solving the BPSP. We have shown that QA outputs a minimum paint swap ratio of $0.329$ at problem size $n=40$ but loses accuracy for larger problem sizes.

Similarly, we numerically estimated the performance of logarithmic depth QAOA via extrapolation and estimate that it has an average paint swap ratio of $0.295$. However, given that there is no quantum advantage for logarithmic depth QAOA when compared to AMP algorithms for sparse optimization problems such as the BPSP \cite{chen2023localalgorithmsfailurelogdepth}, this implies the existence of a suitably applied AMP algorithm that would achieve comparable, if not significantly better performance.

We have provided numerical evidence that the MF-AOA is able to outperform all known classical heuristics, QA and logarithmic depth applications of the QAOA for the BPSP, yielding a paint swap ratio of approximately $0.280$. While we conjecture that AMP algorithms are optimal with respect to the BPSP, it is of further work to provide a rigorous proof of the performance of AMP algorithms for the BPSP.

Given that AMP algorithms seem to work well for the BPSP and unconstrained optimization problems in general \cite{chen2023localalgorithmsfailurelogdepth}, it would be of further interest to study whether such algorithms work in more generalized constrained optimization problems, such as the multi-car paint shop problem.

Lastly, while there is no quantum advantage for logarithmic depth QAOA, it does not rule out an advantage at higher depth (e.g.\ linear depth) or via hybrid solutions. One possible extension would be to identify instances where the MF-AOA fails to find the optimal solution and the depth required by the QAOA to outperform the MF-AOA in expectation value.

\section*{Acknowledgments}

This project was made possible by the DLR Quantum Computing Initiative and the Federal Ministry for Research, Technology and Space; qci.dlr.de/projects/quanticom.
M.G. was funded by the DLR Quantum Computing Initiative through the Quantum Fellowship Programme.
M.G. thanks Ieva {\v C}epait\.e for bringing up the tree-tensor network method that led to the discovery of \cite{QAOA_treetensor}, Kunal Marwaha for clarifying the results of logarithmic depth limitations of the QAOA, David Gross for insightful discussions, and Michael Streif for clarifying the results of their paper. 
The authors gratefully acknowledge the scientific support and HPC resources provided by the DLR. The HPC system CARO is partially funded by Ministry of Science and Culture of Lower Saxony and Federal Ministry for Research, Technology and Space.
The authors gratefully acknowledge the J\"ulich Supercomputing Centre (https://www.fz-juelich.de/ias/jsc) by providing computing time on the D-Wave Advantage$^{\mathrm{TM}}$ System JUPSI through the Jülich UNified Infrastructure for Quantum computing (JUNIQ).

\section*{Author Contributions}

\textbf{M.G.}: Conceptualization, methodology, formal analysis, and primary writing of the manuscript. \textbf{L.C.P.d.S.} and \textbf{T.M.}: Writing -- review \& editing, and equal supporting contributions to the analysis and simulations.
\textbf{M.S.}: Project supervision and funding acquisition.
\textbf{AI Assistance}: Generative AI was used for minor text editing and code generation/formatting. All outputs were independently audited and verified by the authors.

\bibliographystyle{quantum}
\bibliography{bibliography}

\clearpage
\appendix
\section{Angles used in QAOA analysis}
\label{sec:angles}

We include the angles used to evaluate the performance of the QAOA for the binary paint shop problem in the limit $n\rightarrow\infty$. Note that $\bs{\gamma}$ was used with the scaled unitary operator $\exp{-i\gamma \frac{1}{\sqrt{D}}\sum Z_jZ_k}$ for a $(D+1)$-regular graph instead of the usual QAOA unitary operator $\exp{-i\gamma \frac{1}{2}\sum Z_jZ_k}$.

\newcommand{\apptablecaption}[2]{%
  \refstepcounter{table}%
  \par\smallskip
  \noindent\textbf{TABLE \thetable.}\ #1%
  \label{#2}%
  \par\smallskip
}
\newcommand{\appfigurecaption}[2]{%
  \refstepcounter{figure}%
  \par\smallskip
  \noindent\textbf{FIG. \thefigure.}\ #1%
  \label{#2}%
  \par\smallskip
}

\noindent
\begin{minipage}[t]{0.485\textwidth}
\vspace{0pt}

\apptablecaption{Comparison between the large-girth QAOA optimization and
the QAOA simulations of Streif \emph{et al.}~
\cite{QAOA_BPSP}. The simulation data are only available up to
depth $p=7$.}{tab:comparison}

\centering
\footnotesize
\setlength{\tabcolsep}{3pt}
\renewcommand{\arraystretch}{1.0}
\begin{tabular}{ccc}
\hline\hline
Depth $p$ &
$1 - (( 2/ \sqrt{3} )v_p(3,\gamma,\beta)_{\mathrm{best}})$ &
QAOA simulation \\
\hline
1  & 0.675 & 0.675 \\
2  & 0.568 & 0.568 \\
3  & 0.503 & 0.503 \\
4  & 0.462 & 0.462 \\
5  & 0.432 & 0.432 \\
6  & 0.410 & 0.411 \\
7  & 0.393 & 0.393 \\
8  & 0.381 & --- \\
9  & 0.371 & --- \\
10 & 0.363 & --- \\
11 & 0.357 & --- \\
12 & 0.351 & --- \\
13 & 0.347 & --- \\
14 & 0.343 & --- \\
15 & 0.339 & --- \\
16 & 0.336 & --- \\
17 & 0.334 & --- \\
\hline\hline
\end{tabular}
\end{minipage}%
\hfill
\vspace{0pt}

\begin{minipage}[t]{0.485\textwidth}
\apptablecaption{Optimised parameters that can be found in
Refs.~\cite{QAOA_BPSP,Wybo} and converted to the convention used in
this paper.}{tab:bpsp_angles_optimized}

\centering
\footnotesize
\setlength{\tabcolsep}{3pt}
\renewcommand{\arraystretch}{1.0}
\begin{tabular}{@{}c p{0.88\columnwidth}@{}}
\hline\hline
$p$ & \raggedright Optimized angles\par \tabularnewline
\hline
1 & \raggedright $\gamma=(0.4535)$\newline $\beta=(0.3927)$\par \tabularnewline
2 & \raggedright $\gamma=(0.3532,\allowbreak\, 0.6406)$\newline $\beta=(0.5341,\allowbreak\, 0.283)$\par \tabularnewline
3 & \raggedright $\gamma=(0.307,\allowbreak\, 0.5641,\allowbreak\, 0.6532)$\newline $\beta=(0.5879,\allowbreak\, 0.4232,\allowbreak\, 0.223)$\par \tabularnewline
4 & \raggedright $\gamma=(0.2728,\allowbreak\, 0.5089,\allowbreak\, 0.583,\allowbreak\, 0.6679)$\newline $\beta=(0.605,\allowbreak\, 0.4778,\allowbreak\, 0.3613,\allowbreak\, 0.1875)$\par \tabularnewline
5 & \raggedright $\gamma=(0.2519,\allowbreak\, 0.4735,\allowbreak\, 0.5225,\allowbreak\, 0.5951,\allowbreak\, 0.6793)$\newline $\beta=(0.6225,\allowbreak\, 0.5051,\allowbreak\, 0.4167,\allowbreak\, 0.3253,\allowbreak\, 0.1628)$\par \tabularnewline
6 & \raggedright $\gamma=(0.2327,\allowbreak\, 0.4441,\allowbreak\, 0.4881,\allowbreak\, 0.5318,\allowbreak\, 0.6025,\allowbreak\, 0.6813)$\newline $\beta=(0.6293,\allowbreak\, 0.5232,\allowbreak\, 0.4528,\allowbreak\, 0.3883,\allowbreak\, 0.2981,\allowbreak\, 0.1459)$\par \tabularnewline
7 & \raggedright $\gamma=(0.2198,\allowbreak\, 0.4235,\allowbreak\, 0.4605,\allowbreak\, 0.4986,\allowbreak\, 0.5382,\allowbreak\, 0.6043,\allowbreak\, 0.683)$\newline $\beta=(0.6378,\allowbreak\, 0.5327,\allowbreak\, 0.4719,\allowbreak\, 0.4325,\allowbreak\, 0.3632,\allowbreak\, 0.2778,\allowbreak\, 0.1339)$\par \tabularnewline
8 & \raggedright $\gamma=(0.2083,\allowbreak\, 0.4062,\allowbreak\, 0.4428,\allowbreak\, 0.4746,\allowbreak\, 0.5067,\allowbreak\, 0.5419,\allowbreak\, 0.6254,\allowbreak\, 0.7222)$\newline $\beta=(0.6405,\allowbreak\, 0.5385,\allowbreak\, 0.4817,\allowbreak\, 0.4526,\allowbreak\, 0.4101,\allowbreak\, 0.3452,\allowbreak\, 0.2605,\allowbreak\, 0.1198)$\par \tabularnewline
\hline\hline
\end{tabular}

\end{minipage}%
\hfill
\begin{minipage}[t]{0.485\textwidth}
\vspace{0pt}

\apptablecaption{Extrapolated QAOA parameter schedules for the BPSP
large-girth recursion at depths $p=9,\ldots,17$. The schedules were obtained
by extrapolating the smooth optimized lower-depth parameters and were evaluated
without further optimization.}{tab:bpsp_angles_extrapolated}

\centering
\footnotesize
\setlength{\tabcolsep}{3pt}
\renewcommand{\arraystretch}{0.98}
\begin{tabular}{@{}c p{0.88\columnwidth}@{}}
\hline\hline
$p$ & \raggedright Extrapolated angles\par \tabularnewline
\hline
9 & \raggedright $\gamma=(0.1995,\allowbreak\, 0.3894,\allowbreak\, 0.4231,\allowbreak\, 0.4558,\allowbreak\, 0.4779,\allowbreak\, 0.5078,\allowbreak\, 0.5514,\allowbreak\, 0.632,\allowbreak\, 0.7228)$\newline $\beta=(0.6452,\allowbreak\, 0.5424,\allowbreak\, 0.4901,\allowbreak\, 0.4644,\allowbreak\, 0.4298,\allowbreak\, 0.384,\allowbreak\, 0.3201,\allowbreak\, 0.2327,\allowbreak\, 0.1086)$\par \tabularnewline
10 & \raggedright $\gamma=(0.1919,\allowbreak\, 0.3758,\allowbreak\, 0.4073,\allowbreak\, 0.4412,\allowbreak\, 0.4566,\allowbreak\, 0.4831,\allowbreak\, 0.5092,\allowbreak\, 0.5611,\allowbreak\, 0.6379,\allowbreak\, 0.7232)$\newline $\beta=(0.6492,\allowbreak\, 0.5457,\allowbreak\, 0.4962,\allowbreak\, 0.473,\allowbreak\, 0.4428,\allowbreak\, 0.4102,\allowbreak\, 0.3605,\allowbreak\, 0.2982,\allowbreak\, 0.2105,\allowbreak\, 0.0994)$\par \tabularnewline
11 & \raggedright $\gamma=(0.1854,\allowbreak\, 0.3642,\allowbreak\, 0.3941,\allowbreak\, 0.4292,\allowbreak\, 0.4413,\allowbreak\, 0.4612,\allowbreak\, 0.4844,\allowbreak\, 0.5124,\allowbreak\, 0.5696,\allowbreak\, 0.643,\allowbreak\, 0.7237)$\newline $\beta=(0.6527,\allowbreak\, 0.5479,\allowbreak\, 0.5009,\allowbreak\, 0.4788,\allowbreak\, 0.4529,\allowbreak\, 0.4261,\allowbreak\, 0.3894,\allowbreak\, 0.3392,\allowbreak\, 0.2788,\allowbreak\, 0.1921,\allowbreak\, 0.0917)$\par \tabularnewline
12 & \raggedright $\gamma=(0.1798,\allowbreak\, 0.3541,\allowbreak\, 0.3831,\allowbreak\, 0.4177,\allowbreak\, 0.4296,\allowbreak\, 0.4443,\allowbreak\, 0.4641,\allowbreak\, 0.4867,\allowbreak\, 0.5183,\allowbreak\, 0.5776,\allowbreak\, 0.6477,\allowbreak\, 0.7241)$\newline $\beta=(0.6553,\allowbreak\, 0.5497,\allowbreak\, 0.5044,\allowbreak\, 0.4838,\allowbreak\, 0.4602,\allowbreak\, 0.4374,\allowbreak\, 0.4087,\allowbreak\, 0.3692,\allowbreak\, 0.32,\allowbreak\, 0.2615,\allowbreak\, 0.1767,\allowbreak\, 0.085)$\par \tabularnewline
13 & \raggedright $\gamma=(0.1748,\allowbreak\, 0.3453,\allowbreak\, 0.3735,\allowbreak\, 0.4067,\allowbreak\, 0.4197,\allowbreak\, 0.4314,\allowbreak\, 0.4465,\allowbreak\, 0.4678,\allowbreak\, 0.4893,\allowbreak\, 0.5265,\allowbreak\, 0.5848,\allowbreak\, 0.6518,\allowbreak\, 0.7246)$\newline $\beta=(0.6576,\allowbreak\, 0.551,\allowbreak\, 0.507,\allowbreak\, 0.4885,\allowbreak\, 0.4656,\allowbreak\, 0.4459,\allowbreak\, 0.4219,\allowbreak\, 0.3916,\allowbreak\, 0.3508,\allowbreak\, 0.3026,\allowbreak\, 0.246,\allowbreak\, 0.1635,\allowbreak\, 0.0792)$\par \tabularnewline
14 & \raggedright $\gamma=(0.1704,\allowbreak\, 0.3375,\allowbreak\, 0.3652,\allowbreak\, 0.3971,\allowbreak\, 0.4114,\allowbreak\, 0.4216,\allowbreak\, 0.4325,\allowbreak\, 0.4505,\allowbreak\, 0.4695,\allowbreak\, 0.4925,\allowbreak\, 0.5341,\allowbreak\, 0.5915,\allowbreak\, 0.6554,\allowbreak\, 0.725)$\newline $\beta=(0.6596,\allowbreak\, 0.5521,\allowbreak\, 0.5092,\allowbreak\, 0.4922,\allowbreak\, 0.4696,\allowbreak\, 0.4528,\allowbreak\, 0.4317,\allowbreak\, 0.4069,\allowbreak\, 0.3744,\allowbreak\, 0.3337,\allowbreak\, 0.287,\allowbreak\, 0.2322,\allowbreak\, 0.1522,\allowbreak\, 0.0742)$\par \tabularnewline
15 & \raggedright $\gamma=(0.1663,\allowbreak\, 0.3305,\allowbreak\, 0.3579,\allowbreak\, 0.3889,\allowbreak\, 0.4042,\allowbreak\, 0.4133,\allowbreak\, 0.421,\allowbreak\, 0.436,\allowbreak\, 0.4541,\allowbreak\, 0.4722,\allowbreak\, 0.496,\allowbreak\, 0.5413,\allowbreak\, 0.5975,\allowbreak\, 0.6586,\allowbreak\, 0.7255)$\newline $\beta=(0.6611,\allowbreak\, 0.5528,\allowbreak\, 0.5109,\allowbreak\, 0.495,\allowbreak\, 0.4725,\allowbreak\, 0.458,\allowbreak\, 0.4387,\allowbreak\, 0.4182,\allowbreak\, 0.3925,\allowbreak\, 0.3581,\allowbreak\, 0.3182,\allowbreak\, 0.2728,\allowbreak\, 0.2197,\allowbreak\, 0.1423,\allowbreak\, 0.0698)$\par \tabularnewline
16 & \raggedright $\gamma=(0.162654,\allowbreak\, 0.324232,\allowbreak\, 0.351136,\allowbreak\, 0.381669,\allowbreak\, 0.395835,\allowbreak\, 0.406213,\allowbreak\, 0.412487,\allowbreak\, 0.424137,\allowbreak\, 0.438958,\allowbreak\, 0.456281,\allowbreak\, 0.475008,\allowbreak\, 0.502973,\allowbreak\, 0.548089,\allowbreak\, 0.602643,\allowbreak\, 0.661563,\allowbreak\, 0.725815)$\newline $\beta=(0.662575,\allowbreak\, 0.553437,\allowbreak\, 0.511717,\allowbreak\, 0.497215,\allowbreak\, 0.476082,\allowbreak\, 0.461855,\allowbreak\, 0.445043,\allowbreak\, 0.426652,\allowbreak\, 0.404933,\allowbreak\, 0.377068,\allowbreak\, 0.342982,\allowbreak\, 0.303915,\allowbreak\, 0.259659,\allowbreak\, 0.205756,\allowbreak\, 0.133645,\allowbreak\, 0.065832)$\par \tabularnewline
17 & \raggedright $\gamma=(0.1593,\allowbreak\, 0.3186,\allowbreak\, 0.34495,\allowbreak\, 0.37495,\allowbreak\, 0.38855,\allowbreak\, 0.39825,\allowbreak\, 0.4041,\allowbreak\, 0.4147,\allowbreak\, 0.4282,\allowbreak\, 0.4445,\allowbreak\, 0.4622,\allowbreak\, 0.4846,\allowbreak\, 0.5158,\allowbreak\, 0.5605,\allowbreak\, 0.6139,\allowbreak\, 0.666,\allowbreak\, 0.7261)$\newline $\beta=(0.6639,\allowbreak\, 0.554,\allowbreak\, 0.5122,\allowbreak\, 0.4991,\allowbreak\, 0.4791,\allowbreak\, 0.4656,\allowbreak\, 0.4499,\allowbreak\, 0.4328,\allowbreak\, 0.4134,\allowbreak\, 0.3896,\allowbreak\, 0.3605,\allowbreak\, 0.3263,\allowbreak\, 0.2876,\allowbreak\, 0.2448,\allowbreak\, 0.194,\allowbreak\, 0.1262,\allowbreak\, 0.0624)$\par \tabularnewline
\hline\hline
\end{tabular}

\end{minipage}%

\vspace{0.55\baselineskip}

\noindent
\begin{minipage}[t]{0.485\textwidth}
\vspace{0pt}
\centering
\includegraphics[width=\linewidth]{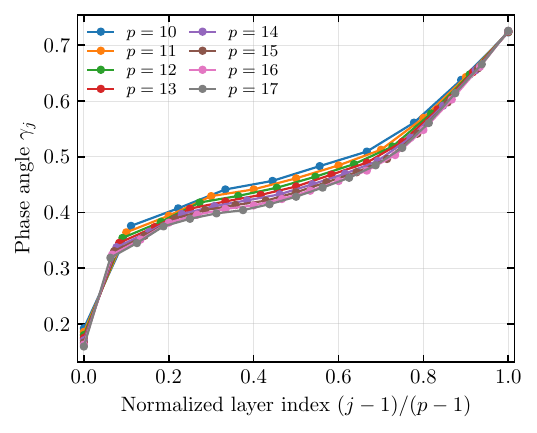}
\vspace{-0.45\baselineskip}

\appfigurecaption{Normalized QAOA parameter schedules used in the large-girth BPSP calculation for depths $p=10,\ldots,17$. The phase-separation angles $\gamma_j$ and mixer angles $\beta_j$ are shown as functions of the normalized layer coordinate $s_j=(j-1)/(p-1)$. The collapse of the curves with increasing $p$ demonstrates the smooth depth dependence of the parameter schedules. The schedules for $p\geq9$ were generated by extrapolating the optimized lower-depth schedules and subsequently evaluated without further optimization. The phase angles are quoted in the convention $\exp[-i\gamma D^{-1/2}\sum_{\langle ij\rangle}Z_iZ_j]$ with $D=3$. Plotted values from Table~\ref{tab:bpsp_angles_extrapolated}.}{fig:gammas}

\end{minipage}
\hfill
\begin{minipage}[t]{0.485\textwidth}
\vspace{0pt}
\centering
\includegraphics[width=\linewidth]{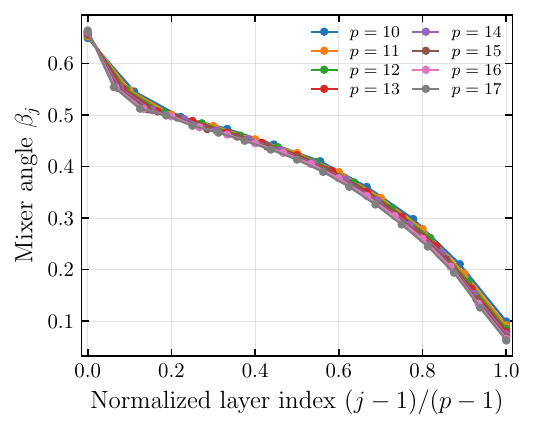}
\vspace{-0.45\baselineskip}

\appfigurecaption{See Fig.~\ref{fig:gammas}.}{fig:betas}

\end{minipage}

\end{document}